# QUASARS AND LARGE SCALE STRUCTURE OF THE UNIVERSE

V.N.Lukash

Astro Space Centre of Lebedev Physical Institute,
Profsoyuznaya 84/32 , 117810 Moscow, Russia

The majority of bright distant quasars ($z \gtrsim 1$) may form in massive mergers appearing in compact galaxy groups in/and young clusters. The expected tests are (i) large correlation signal for medium-$z$ QSOs ($1 < z < 2$) and (ii) direct search for quasar groups (QGs) indicating positions of distant pre-superclusters which later will evolve to the "systems" like the local Great Attractor or Shapley concentration. We discuss large QGs with more than ten members within regions $\leq l_{LS} \sim 100 - 150\, h^{-1} Mpc$, tracing the enhanced density regions at $z \lesssim 2$. These early large scale structures (i) provide a natural way to "bias" the distribution of Abell clusters, and (ii) suggest that the spectrum of primordial density perturbations is nearly flat at scales encompassing both cluster and GAs, $l = \pi k^{-1} \in (10, 100) h^{-1} Mpc : \Delta_k^2 \sim k^3 P(k) \sim k^\gamma, \gamma = 1^{+0.6}_{-0.4}$.

Current discussion on the cosmological models of large scale structure (LSS) formation is strongly stimulated by recent observational progress. Today, we have few independent evidences for existence of large scale structures up to a typical scale $l_{LS} \cong 100 - 150 h^{-1} Mpc$. The most important data come from the Great Attractor ($z \leq 0.03$), distributions of clusters of galaxies ($z \leq 0.1$) and pencil beam galactic surveys ($z \leq 0.3$). It is now on agenda to test ever deeper tracers, e.g., radiogalaxies and active galactic nuclei (AGNs), quasars (QSOs), absorption ($L_\alpha, CIV$) and emission (21 cm) clouds, X-ray clusters and others, to make sure how far these large structures extend in the past. The result would allow direct dynamical reconstruction of the primordial spectrum of density perturbations within scales $10 - 100 h^{-1} Mpc$. Together with $\Delta T/T (\Theta \geq 1^o)$ measurements this could certainly solve the problem of the primordial spectrum in the whole range of large scales.

Here, let me concentrate on the relationship between quasar groups and enhanced density regions that are called distant Great Attractors (GAs)[1]. As for the local GA, by distant GA we understand a patch of enhanced total density, scaled in any dimension from larger than richest cluster size and up to $\sim l_{LS}$ (and further, if any). The opposite construction is void (Great Repulsor), a patch of decreased total density. Both notions - GAs and voids - are quasilinear systems still expanding with the Universe (the density variation may consist $\sim 10 - 30\%$ on scale $\sim 100 h^{-1} Mpc$) which differ them drastically from the objects just collapsed or collapsing at least in one direction – galactic clusters, filaments and walls – the latter direction size being $\lesssim 10 h^{-1} Mpc$.

There are two principal steps on the way to relate distant QGs and GAs:

∗ Clusters, where they are concentrated (the superclusters), trace the mass density enhancements;

∗ Clusters of distant clusters contain (at least some) QGs.

The first point is confined today in the nearby region where peculiar velocity measurements are done, and by the alignment of the cluster dipole with microwave background dipole . As for the second point, there are few observational indications (I stress here only two of them):

1) The QSO correlation functions at different redshifts and comparison with clusters' and galaxies' correlation functions.

2) Direct investigations of the environments of nearby QSOs.

The QSO correlation function displaying strong QSO clustering at $r \leq 10 h^{-1} Mpc$, a weaker clumping at $\sim 20 - 100 h^{-1} Mpc$ and the absence of correlations for $r > 150 h^{-1} Mpc$, evolves explicitly growing to smaller redshifts: for $z \Rightarrow 0$ it looks like the distribution of galaxies whereas for $z > 2$ the QSOs seem to be randomly distributed. According to [2,3], $\xi_{qq} \sim (1+z)^{-a}$, with $a \sim 2-3$. Having analysed 8 homogeneous surveys of QSOs we found that the main contribution to $q$-$q$ correlation signal comes from medium-$z$ quasars with the correlation radius $\sim 10\,h^{-1}\,Mpc$ $(1 < z < 2)$[4]. When approximating it by $z = 0$ due to the linear perturbation law ($\xi_{qq} \sim (1+z)^{-2}$), the correlation function for those QSOs would correspond by now to that of the reach clusters of galaxies. As the nearby QSOs ($z < 1$) are distributed much more like galaxies than like clusters, they seem to be a different population which is more closely connected to poor environments like loose groups of galaxies (whereas the bright distant quasars mostly form in compact groups in preclusters). One of the principal tests here could be a clear detection of the decay of $\xi_{qq}$ beyond $z > 2$.

Today's observations display that QSOs at $z > 0.5$ are frequently associated with rich clusters and compact groups, while those at $z < 0.5$ seem to be found in sparse poor systems like loose galactic groups. It gives us a guess that quasars may form in merging and intercting galaxies originating in galaxy concentrations (the density enhancements). These merging effects generate instability processes in host galaxies resulting in the gas infall and accreation material supply (to a massive black hole) enough for the QSO burning. Regarding bright distant quasars ($z \gtrsim 1$) such conditions can really exist in young clusters which are in process of the first contraflows' origin, i.e., well before the cluster virialization and X-ray gas appearing. Taking into account that the first violent crossings (caustics of LSS forming) of the cosmic primordial medium must certainly arrise in the central regions of young preclusters at $z \sim 2.5$, we may relate this epoch with the quasar peak existing at redshifts $z \sim 2-3$. Of course, not every QSO we do associate with collapsing compact groups in young clusters (there were few generations of QSOs depending on the environmental physical conditions which provided due to merging activity for the galactic matter infall and formation of the accreating gas disk). The smaller masses collapsed earlier and could not ensure intensive galaxy interactions, while the larger LSS masses collapsed later in the medium with a much smaller fraction of the gas (transfering to stars) already totally ionized.

If so, then at least two topics are on the agenda:

* It is possible to find distant GAs by groups (and even pairs) of QSOs in scale of few tens of Mpc ($l_{GA} \lesssim 100 h^{-1} Mpc$).

* Relationship between the epochs of first generation of caustics in preclusters ($z \sim 2-3$) and in GAs ($z \sim 0$) prompts the true model of LSS formation.

The QSO test for GA search proposed in the first topic can be successfully used at $0.5 \leq z \leq 3.5$, where large QSO number densities are observed. Today we have information on eleven QGs with more than ten members in a group and the number density enhancement (relative to the background QSO density) more than factor of two [5]. The most famous in the list is the group at $z = 1.1$ consisting of 25 QSOs within $\sim 60h^{-1}Mpc$[6]. All QGs are found for $z < 2$ and not seen at larger redshifts (though the spatial number density of QSOs grows sharply up to $z \sim 2.5$), the typical QG-sizes are about supercluster scales ($\sim l_{LS}$).

If the formation epochs of clusters and GAs are really "separated" by the interval $\Delta z \sim 2.5$ ("subcluster formation" proceeds in a broad interval centered at $z^*_{CL} \sim 2.5$, while "cluster formation" goes on nearly now) we have the following simple estimate for the spectrum of Gaussian density perturbations $\delta \equiv \delta\rho/\rho$ within the dynamical scale range $l = \pi k^{-1} \in 10 - 100 h^{-1} Mpc$:

$$\Delta_k^2 \sim k^\gamma, \quad \gamma = 1^{+0.6}_{-0.4} \tag{1}$$

$$\delta^2 = (1+z)^{-2} \int \frac{dk}{k} \Delta_k^2,$$

where $\delta^2$ is the variance. It predicts the correlation function of the dynamical mass $\xi(r) \sim r^{-\gamma}$, which means that clusters are clumped in regions overdensed by GA-scale perturbations. By now, the perturbations on supercluster scale seem to be clearly outlined (according to eq.(1) $\delta_{GA} \sim \delta_{LS} \sim 0.2$) when the dynamical scale ($\delta_D \sim \delta_{Cl} \sim 1$) is closed to $10h^{-1}Mpc$ (which is the scale of largest clusters and the width of the walls and filaments of LSS network).

Such a flat spectra bring about the following conclusions.

(i) Clusters today should be most developed and concentrated in central parts of GAs (clusters trace the mass in dense regions), while sparse young clusters may be found nearby or inside voids.

(ii) If the majority of bright QSOs broadly peaked at $z \sim 2-3$ is formed in preclusters in GA-locations then the mean separation between QSOs at $z \sim 2.5$ should correlate with $l_{LS}$ which is actually in good agreement with observations.

(iii) The Gaussian flat spectra (1) create by $z < 1$ a great variety of coherent structures (the hierarchies of walls, filaments, voids and GAs) produced by broadscale (in $l \in 10 - 100h^{-1}Mpc$) perturbations with close initial amplitudes.

(iv) Since galaxies form before clusters, their first generation is not modulated by GA-perturbations. However, merging and generating processes for galaxies going most active in rich environments (the dense regions), lead to the successive generation of large galaxies (which give birth to bright QSOs) namely at GA locations. The test could be a search of dwarf galaxies in voids.

(v) Obviously, spectra (1) are more flat than that in standard CDM (the latter anticipates $\Delta_k^2 \sim k^{2-3}$ in the scale range). As the COBE $\Delta T/T$ detection indicate consistency with Zeldovich-Harrison (ZH) spectrum at very large scale ($\Delta_k^2 \sim k^4, l \sim 1000h^{-1}Mpc$) then the turn from the flat part (1) to ZH asymptotic should happen at supercluster scale $\sim 150h^{-1}Mpc$, which can obviously be the real feature of the primordial spectrum (contrary to the case of galaxy clusters to be the consequence of current dynamical time).

Summarising, we can say that available data back the following conjectures.

∗ QGs may be good tracers of enhanced matter density regions (distant GAs), they render in sizes up to $l_{GA} \sim 100 - 150h^{-1}Mpc$ and extend by $z_{LS} \sim 2$.

∗ Majority of distant QSOs ($1 \lesssim z \lesssim 3$) can be created in massive merges forming in collapsing compact groups in precluster clumps.

Both points are selfconsistent if and only if QSOs belonging to QGs were born not far before the time where given QG is observed, which is also consistent with a relatively short lifetime of medium and bright QSOs. It means that very distant QSOs at $z > z_{LS}$ appear more or less random in space (GA perturbations are small yet). On the contrary, the medium-z QSOs stemming at $z < z_{LS}$ originate already in groups (the formation time of galaxy clusters is strongly modulated by GA-scale perturbations). Note in this connection that late QSOs ($z < 1$) form in poor environments as their formation is delayed. The evolutionary epoch of compact groups is mainly over by now, so, the remnants of the distant QSOs may be hidden in large E-galaxies, while the collapsing time of loose groups and young clusters today is comparable with the Hubble time.

Certainly, both points lead to obvious conclusion that the part of the primordial spectrum between the dynamical scale ($\sim 10h^{-1}Mpc$) and the coherence length ($\sim 100 - 150h^{-1}Mpc$) is nearly flat (see eq. (1)).

This work has been supported in part by Russian Foundation for Fundamental Research (code 93-02-2929) and International Science Foundation (code MEZ300).